# A novel set of algorithms to recognize galleries of ambrosia beetle in computerized axial tomography of trees trunks

Andres E. Dolinko[1,4], Yasmil Costales[2], Cecilia Carmarán[1] and Esteban Ceriani-Nakamurakare[3,4]

[1] CONICET—Universidad de Buenos Aires, Instituto de Micología y Botánica, Argentina
[2] Universidad Favaloro, Buenos Aires, Argentina
[3] CONICET – Universidad de Luján, Departamento de Tecnología, Laboratorio de Fitopatología, Argentina
[4] Universidad de Buenos Aires, Facultad de Agronomía, Departamento de Ingeniería Agrícola y Uso de la Tierra, Cátedra de Física, Buenos Aires, Argentina

**Abstract**

*Megaplatypus mutatus* is an ambrosia beetle that attacks several species of trees by making galleries in the trunks where its larvae and associated fungi develop. This damage spoils the wood for commercial use and cause stem breakage in front of strong winds. Due to the insect's cryptic lifestyle, gallery analyses have usually been studied by destructive methods. However, they alter the homeostasis of the insect-fungi interaction, modifies the topology of the gallery and, more importantly, does not reveal the 100% complex structure made by the insect. Therefore, a novel way to study this structure is by imaging the galleries by means of computerized axial tomography (CT). This method allows obtaining a three-dimensional representation of the gallery and the pupal chambers to be studied, while the wood and insect sample is not disturbed and generates a high amount of data. The isolation of the galleries and pupal chambers from the CT background images is not simple, because there is not enough contrast between the grey levels of the galleries and the marks generated by internal components of the trunk itself. In this paper, we present a robust algorithm that allows automating the isolation of the galleries and pupal chambers from CT trunk images which can be used in a broad spectrum of image analysis.

Corresponding author: adolinko@df.uba.ar

**Introduction**

Platypodid ambrosia beetles (Platypodinae) are an important group of weevils found in forest ecosystems that usually are among the first wood-degrading agents arriving on weakened or dead trees consumption (Vanderpool et al. 2017, Ceriani-Nakamurakare 2022). However, *Megaplatypus mutatus* (Curculionidae: Platypodinae) is an ambrosia beetle native to South America that causes damage to vigorous trees. This species exhibits generalist habits and is considered a key pest that threatens the expanding forest industry in countries such as Argentina, Brazil, Colombia, Italy, Uruguay, Paraguay and Peru (Alfaro et al. 2007). Preferred hosts include widely planted and traded genera such as *Eucalyptus* and *Populus*, but have a broad spectrum of susceptible host tree species. Despite the importance of this insect, gallery studies are practically null, mainly because of the cryptic lifestyle and the complexity involved in the study. Santoro (1963) estimated the length of the gallery manually by analyzing a small number of samples. Most probably, because it is an activity that is highly time-consuming and obtains variable results depending on the observer and the methodology itself, i.e., the possibility of leaving portions of the tunnel undiscovered at the moment of cutting the sample. An approach that circumvents these limitations is the use of non-destructive methods enhanced with image recognition algorithms, thus helping to preserve the sample integrity and reduce processing time and human error to constant and minimal values.

Ceriani-Nakamurakare et al. (2016) made considerable progress in the study of the gallery topology of *M. mutatus* using Computed Axial Tomography (CT). The X-ray CT technique provides slice grey-level images of the sample that correspond its mass density distribution This advanced technique allows imaging the tunnels inside the trunks, thus being able to visualize the three-dimensional structure of the galleries, which facilitates the study of the beetle dynamics, its behavior, as well as the estimation of descriptive parameters of the galleries. However, the isolation (i.e. the separation of the of the structures of interest from the background) of the galleries and pupal chambers alone is not a simple task due to the tomographic images contain the different regions and components forming the trunk, like the growth rings, internal branches, regions of different densities, among others, highlight the complexity of automatic isolation procedure. There is not a clear contrast between the grey level of the galleries and pupal chambers and other regions in the trunk imaged by the tomographic procedure. Depending on the tomographic resolution of the tomograph and the longitude of the trunk section to be studied, the whole set of tomographic slices can be composed of several hundreds of images, making the manual technique highly time consuming. So, the isolation must usually be made by a manual procedure consisting in spotting the galleries in each slices of the CT. Besides, the manual procedure becomes monotonic and can lead to errors causing an imprecise marking of the objects of interest in the tomography. In contrast, a computer algorithm applies the same detection rules to the whole set of images in the tomography, which provides a more reliable result. Because of that, the development of efficient recognizing algorithms results of great interest in the field of biological image processing (Uchida 2013). Biological images have been also used as input data in computer simulations for the analysis of structural color in the field of biomimetics (Dolinko 2009, Dolinko 2021) and the analysis of electrostatics forces in the release of pollen in anemophilous and buzz pollination (Dolinko 2018, Galati 2019).

In this paper, we present a three-step procedure algorithm with minimum intervention of a human operator. The procedure consists of three separated algorithms that use the tomographic digital images of tress trunks as input data to analyse galleries and pupal cameras made by boring insects. The first algorithm is the main algorithm and represents the key part of the procedure. It consists of an image recognition algorithm that sweeps each tomographic slide searching for objects sizes compatible with galleries and pupal chambers. The second algorithm basically performs a postprocessing on the output of the main algorithm making a thresholding. Lastly, the third algorithm identifies among galleries and pupal chambers recognized by the first two algorithms. Finally, this information can be used to visualize the complete structure and be able to quantifies different geometric parameters of the three-dimensional structure composed by the galleries and pupal chambers to charaterize the galleries. In the following, we describe these algorithms and discussed the obtained results with real tomographic sets of affected trunks with damages caused by ambrosia beetle.

**The algorithms**

As mentioned above, the processing of the tomographies is performed by means of three algorithms. The first algorithm recognizes any spot or mark in the image with a characteristic size of the galleries and pupal chambers. In particular, the algorithm searches for spots or marks with the dimensions of the diameter of the gallery or pupal chambers. As the output, this algorithm provides a set of images, each one corresponding with its original tomographic image, but in these output images, the gray levels of the pixels are related with a measure of an abstract Euclidean distance that measures the similarity of the marks in the original image to a certain kernel image with the characteristic size of the gallery or pupal chamber diameter. More precisely, a pixel near-black color (near 0 gray level) indicates that the image in that region has a pattern with a size very near to the characteristic size of the gallery and pupal chamber size. On the contrary, a white color (near 255 gray level) indicates that in that region, there is no mark with the required size. The second algorithm thresholds the output of the first algorithm. Its output is a set of binary images, each one corresponding with its original tomographic image, in which the galleries and pupal chambers have a value of one and any other region have a value of zero. These images are written down to the hard disk to be able to do a final manual cleaning to erase any residuary noise or spurious marks. Lastly, the third algorithm analyses the binary images and recognizes the pixels corresponding to galleries from the pixels corresponding to pupal chambers. This algorithm generates a new group of images where gallery pixels are labeled with a value equal to 1 and pupal chamber pixels are labeled with a value equal to 2.

**Main algorithm: gallery isolation from tomographic images**

The main algorithm is an image recognizer that evaluates the similarity of the size of any spot in the image with the characteristics of the diameter of the galleries and pupal chambers. Since the diameter of galleries and pupal chambers is roughly similar, there is just one characteristic size $d_{ch}$ to be identified (i.e. the pupal chamber or gallery diameter). The procedure begins by creating a small patch image containing a circular spot whose diameter corresponds to the characteristic size $d_{ch}$. From now on, we call this image the *kernel image*. The kernel image should be adjusted for each tomographied trunk since tomography parameters, image amplification and trunk dimensions must vary among different tomographies samples (see Figure 1).

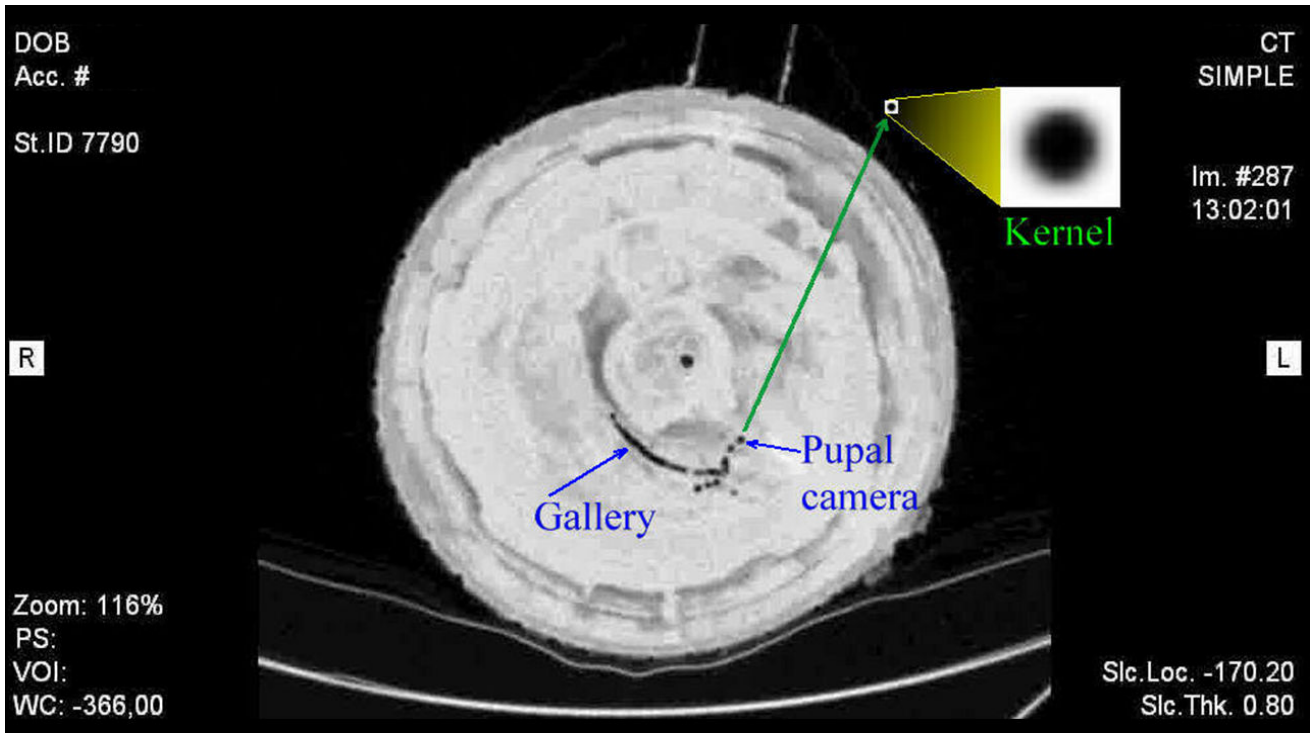

-**Figure 1:** Galleries and pupal chambers in the tomographic image slice (transverse slice of tree trunk) and the generation of the kernel image of characteristic size $d_{ch}$ . -

Euclidean norm between the grey levels of both images along with several diametral cuts of the images at different angles, for searching, if in the image there is at any angle a spot with the characteristic size $d_{ch}$. Therefore, this abstract distance will give a measure of geometric similarity of the images.

The algorithm begins by sweeping each tomographic image. The entire set of images is swept pixel by pixel from left to right and from up to down, getting a patch of the image with the same size as the kernel. Then, the diametral cross-section profile of the patch for each angle from 0 to 90 is compared with the diametral cross-section profile of the kernel at the same angles (see Figure 2).

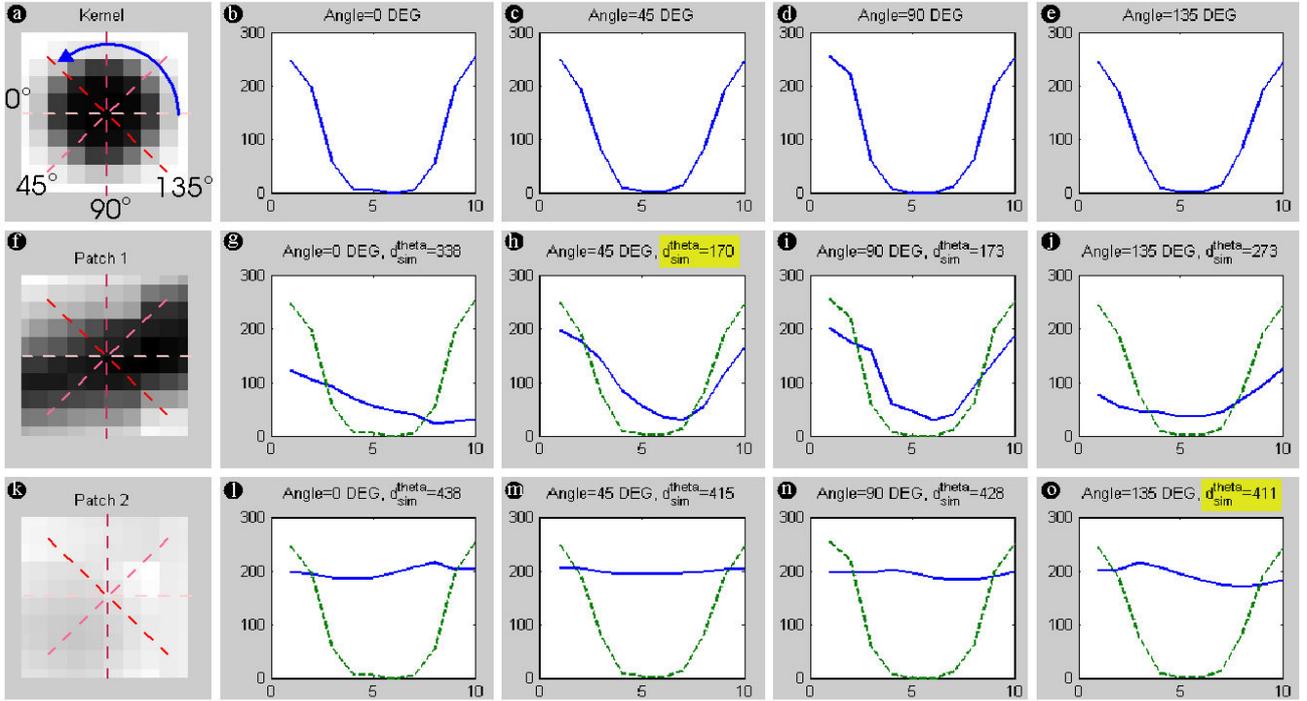

**-Figure 2**: **(a)** The kernel used as reference; **(b-e)** kernel profile at different angles; **(f)** and **(k)** patch image at two different locations of a tomography slice; **(g-j)** and **(l-o)** corresponding profiles for images (b) and (c) at different angles (blue solid line) compared to the kernel profile at the same angle (green dashed line) and their corresponding Euclidean distance. -

Both profiles are compared by using an Euclidean norm defined as

$$d_{sim}\theta = \sqrt{\left(\sum_{i=1}^{N}(Pf^{Im}_i - Pf^{Krn}_i)^2\right)} \qquad (1)$$

where $d_{sim}^{\theta}$ is an abstract euclidean distance, $i$ is the i-th pixel of the profiles of longitude $N$, $Pf^{Im}$ is the profile taken on the patch at an angle $\theta$ and $Pf^{Krn}$ is the profile at the same angle taken on the kernel. Then, $d_{sim}^{\theta}$ quantifies the similarity between two one-dimensional profiles. When the distance is near to 0, this indicate that the profile at that angle is similar to that of the kernel at the angle $\theta$. Then, the minimum distance $d_{sim}^{Min}$ among the set of explored angles $\theta$ is taken and the central pixel of the patch is filled with this value of distance. Lower distances indicates that a gallery cross-section or pupal chamber was detected. For each patch position, the comparison is explored with angles from $\theta = 0$ to 90. Ideally, this should be made in THE step of 1 degree. However, to minimize the computation time of each complete image and serie, it was verified that exploration in steps of 45 degrees gives acceptable detection results. (i.e. just the set $\theta = 0, 45$ and 90 degrees is explored). In this manner, any spot with the characteristic transversal size $d_{ch}$ in some direction in the image is enhanced through of low values of $d_{sim}^{Min}$. Finally, the output of this algorithm is an image of the same size as the original where each pixel has a value of $d_{sim}^{Min}$.

Figure 2 shows an example of the evaluation for two different patches (fig. 2(f) and 2(k)). fig. 2(a) shows the kernel used in this case, while fig 2(b) to 2(e) the diametral kernel profile at different angles, which are shown in different tones of red dashed line in fig 2(a). Figs 2(g-j) and 2(l-o) show the profiles

(solid blue line) for the patches shown in fig. (f) and (k), respectively, compared to the profiles of the kernel (dashed green line) at the same angles. The Euclidean distance $d_{sim}\theta$ at each angle is shown for each angle for both example cases. For each patch, the value of minimum euclidean distance $d_{sim}^{Min}$ is highlighted in yellow. In this manner, the direction in which the corresponding patch adjustS to the kernel profile is detected. In the case of patch in fig. (f), a gallery profile is detected at 45 degreeS. In the case of fig. 2(k), although a minimum value of is obtained, these values are high, indicating that no spot of characteristic size $d_{ch}$ is detected, indicating that the evaluated patch image has no gallery or pupal chamber nor part of them in it.

Finally, a new image called *distance map*, with a distribution of the obtained values $d_{sim}^{Min}$ is generated. This new image, which has the same size of the original tomographic image, is obtained as follows: the value of $d_{sim}^{Min}$ obtained for each patch in the tomographic image is associated with the central position pixel of this patch in the *distance map*. The distance map looks similar to the original tomographic image with the difference that in the distance map the spots corresponding to the galleries and pupal chambers are enhanced in relation to the rest of the image. Any spot with a different characteristic size $d_{ch}$ is attenuated (towards white tones), while any spot in which any diametral size corresponds to $d_{ch}$ is enhanced (towards black tones). As shown in figure 3, it can be observed that the only enhanced spots (darker) in the figure corresponds to the galleries and pupal chambers in the tomographic slice.

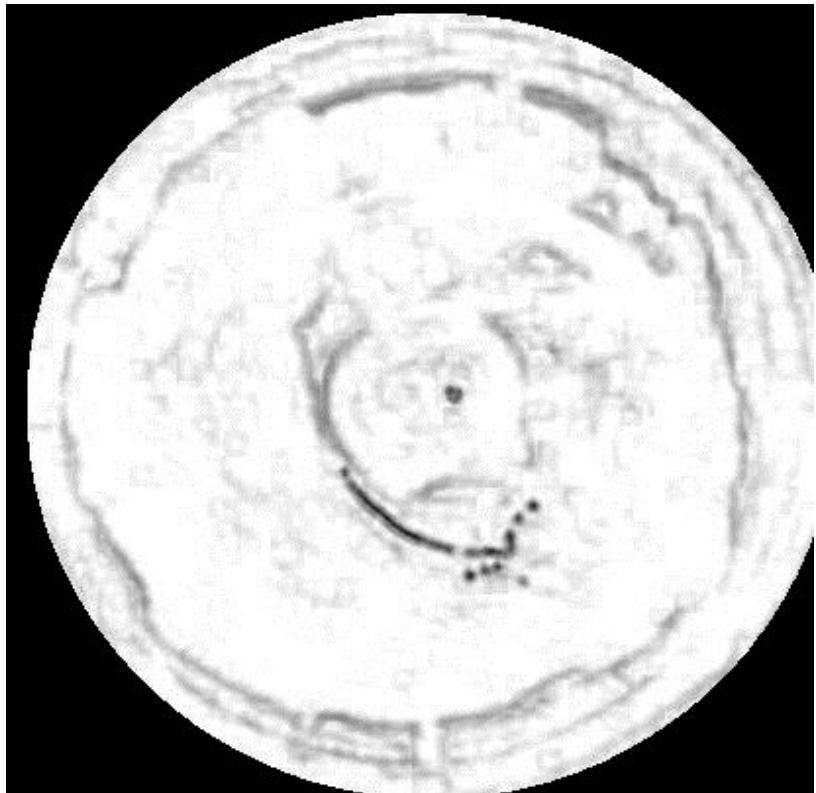

-**Figure 3**: *Distance map* $d_{sim}^{Min}$ resulting from the application of the main algorithm corresponding to the tomographic slice showed in figure 1.

**Second algorithm: thresholding of distance maps**

The objective of the second algorithm is to threshold the image given by the first algorithm and clean residuary noise resulting from erroneously detected spots in the main algorithm. Although the first algorithm enhanceS any spot of characteristic size $d_{ch}$ in the image, any part in the image has a value $d_{sim}^{Min}$, as shown in figure 2, even parts that have no regions with galleries or pupal chambers. Because of that, a selection of adequate values of $d_{sim}^{Min}$ revealing the galleries and pupal chambers must be performed. In order to do this, the whole set of images is swept. First, the output image of the first algorithm IS converted to grey levels where the maximum value of $d_{sim}^{Min}$ is set as 1 and the minimum value is set to 0. Each image is threshold according to a parameter called *thresholdval* introduced manually that can range from 0 to 1 and that selectS the values of $d_{sim}^{Min}$ that are related to a positive detection of galleries and pupal chambers. Therefore, for each image $d_{sim}^{Min}$ a new image is generated where pixel values over *thresholdval* are set to 0 (negative detection) and values under *thresholdval* are set to 1. In this manner, a set of binary images is generated where white values reveal the galleries and pupal chambers and black values are set for the rest of the image area. The adequate value for *thresholdval* ranges from 0.5 to 0.7 and depends on the characteristics of the output images given by the first algorithm for each tomography set.. It is worthwhile mentioning that low values of *thresholdval* lead to low noise but pixels corresponding to galleries and pupal chambers could be erroneously avoided. On the contrary, high values of *thresholdval* will detect the majority of pixels corresponding to galleries and pupal chambers, but will result in a higher level of noise. Therefore, the precise value of *thresholdval* results in a tradeoff between THE level of detection and noise quantity, and is set for each tomography set by means of visual inspection of the output of the second algorithm, as shown in Figure 4 for different values of *thresholdval*. Figure 4(a) shows the thresholding result for *thresholdval* =0.3, giving an under-detection image; i.e. the image appears with no noise but a part of the galleries and pupal chambers does not appear. On the other hand, figure 4(b) shows the thresholding result for *thresholdval* =0.8, giving an over-detection image; i.e. all the galleries and pupal chambers are shown, but the image results still have noise. Finally, figure 4(c) shows the optimum value for this test, given by a *thresholdval* =0.56, where all the set of galleries and pupal chambers appears and almost no noise is present.

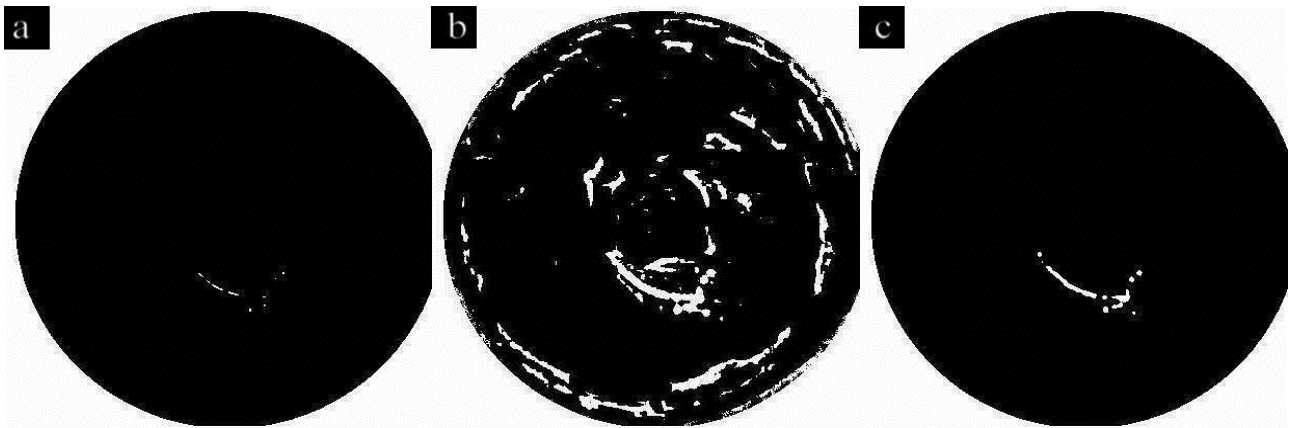

- **Figure 4**: Thresholding of distance map showed in figure 3 for (a) *thresholdval* =0.3; (c) over-detection, with a threshold image with a high value of *thresholdval* = 0.8; and (d) threshold image with the optimum value of *thresholdval*=0.56.-

Finally, a second strategy is applied to cancel residuary noise. Some noise results from the false detection of spots that are proper of the structure of the trunk and coincide with the characteristic size $d_{ch}$ of the galleries and pupal chamber diameters. The particularity of these spots IS that they are approximately repeated along with the whole set of tomographic images of the trunk. To detect these type of residuary spots, their cumulative effect along the whole set of images is detected. To do this, the whole set of output images resulting from the second algorithm IS added up, obtaining an image, called *cumulative map*, where the value of each pixel representS the times a detected spot is repeated along with the whole set of images (see Fig. 5a). Knowing that galleries and pupal chambers extend along a few slices, the pixels with values over certain valueS indicating a repetition along a great number of slices are canceled. The pixels to be eliminated form a mask, which we call *automask* (see Fig. 5b). For example, since the galleries have a diameter of approximately 8 pixels, the pixels representing regions of galleries in the *cumulative map* will have values around this number. For pupal chambers, generally, theY extend along 30 slices, and up and down chambers are generally aligned. Therefore, values for pupal chambers in the *cumulative map* will have values ranging from 25 to less than 70. Then, by applying the automask to the whole set of tomographies, the repetitive noise is canceled (Fig. 5c). Emplying the colorbar, it can be observed in figure 5a that galleries and pupal chambers do not exceed 100 slices. On the contrary, a spot corresponding to the kernel of the trunk has a value over 200. In figure 5b, we have degenerated automask and in figure 5c, the automask was used to cancel the corresponding pixels with repetitive spurious spots.

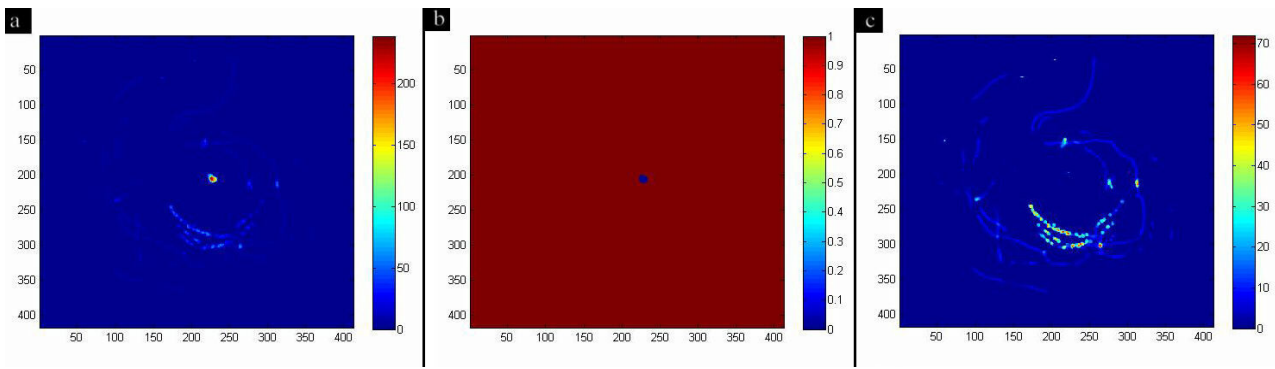

- **Figure 5**: Automask procedure: (a) *cumulative map*; (b) *automask*; (c) filtered *cumulative map* with the obtained *automask*.-

**Third algorithm: separation between galleries and pupal chambers**

The objective of the third algorithm is to distinguish between galleries and pupal chambers in the output binary images of the second algorithm. We propose a very simple algorithm which is based in the information provided by the filtered Z-cumulative map (figure 5c) which approach is similar to that of the *automask* procedure. Observing figure 5c, it can be notice that since pupal chambers are aligned in Z, they cover a larger number of slices than the galleries, which extend in Z along a number of slices corresponding to their own diameter. Then, using this fact, we generate a mask with those pixels exceeding a certain threshold value of repetition along Z. This threshold value corresponds to the maximum diameter of the gallery in the tomography being analyzed, which is called thresholdgal. The algorithm assumes that there will not be enough superposition of galleries in Z in any point on the cumulative map that will surpass the *thresholdgal* value. Therefore, we generate a new mask called *cammask* where pixels exceeding the thresholdgal are marked with 2 while pixels under this value are zero. We call to this images the *camgalmap*. Then, we apply this cammask to the whole set of slice. In this manner, in each slice, the galleries are represented by a 1 value while the pixels belonging to the pupal chambers are represented by a 2 value, allowing an easy way to separate both structures: the galleries and the pupal chambers.

The algorithm described above is efficient and fast. However, it has the particularity that marking the region of the gallery in immediate contact with the pupal chamber as pupal chamber. This particularity is good or not, depending of the definition for the pupal chamber limits in relation to the gallery. That is, if the pupal chamber is defined as to be inserted in the gallery (see figure 6a) or over the gallery (see figure 6b). In the first case, the presented algorithm would result adequate. However, such a definition could affect the accounting for the longitude and volume of the galleries, that it would be if the pupal chamber would not be present. Therefore, the last definition (figure 6b) seems to be more adequate.

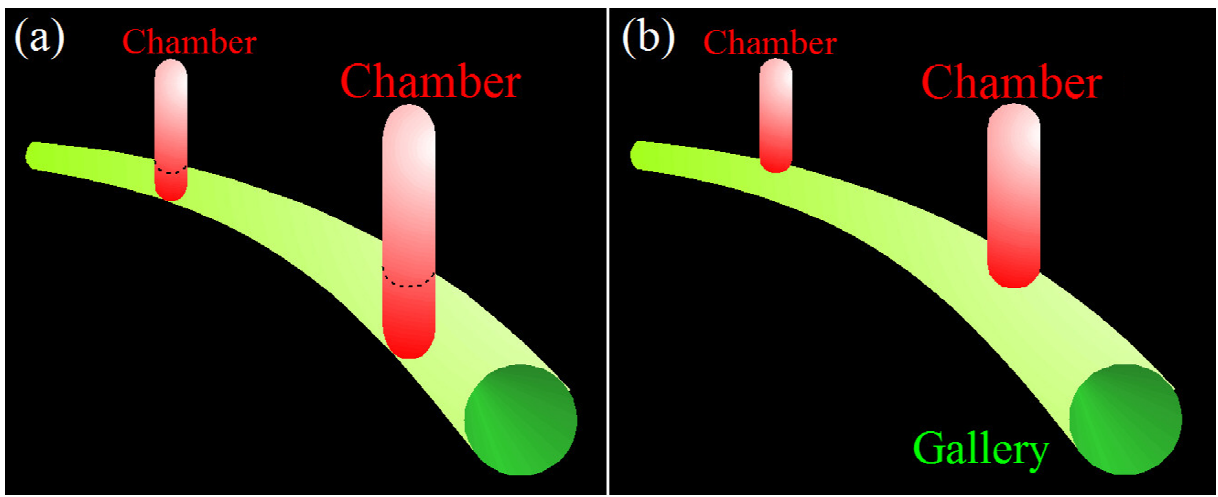

- **Figure 6**: Definitions of limits between gallery and pupal chambers: (a) pupal chamber inserted in the gallery; (b) pupal chamber on the gallery. -

Therefore, taking the definition presented in figure 6b, the results provided by the separation algorithm requires a postprocessing to assign the insertion part of the pupal chamber as part of the galleries (that is, converting pixels labeled as 2 to pixels labeled as 1). In order to do this, we propose a second post-processing algoritm to be applied. The aim of the algorithm is to identify cam pixels surronded by enough gal pixels in each slice, in order to change those cam pixels by a gal pixels. Basically, this algorithm sweeps the *camgalmap* in Z and searches for pixels corresponding to galleries surrounding the pixels corresponding to cams. The algorithm evaluates each slice with a 3x3 patch. When the central pixel in the patch is labeled with 2 (white, i.e. belonging to pupal chamber in our coding), the algoritm count the number of grey pixels (labeled with 1, i.e. indicating gallery regions) in the 8 remaining positions of the patch. If the algorithm counts a sufficiently high number of 1-labeled pixels, it assumes that the 2-labeled pixel is surrounded by gallery and the 2-labeled pixel is replaced by a 1-labeled pixel. In this manner, the postprocessing algorithm replaces those pixels of pupal chambers with pixels of gallery. The optimal number of grey pixels ($n_{grey}$) used to change the central pixel must be calibrated by test and error. If $n_{grey}$ =8, the entire section of the pupal chamber will not never be detected. On the other hand, if $n_{grey}$ =1, due to a any noisy grey-pixel (false galery part) would change a pupal-cam pixel that shouldn't be changed. We obtained an ideal value of $n_{grey}$ =4 (see figure 7).

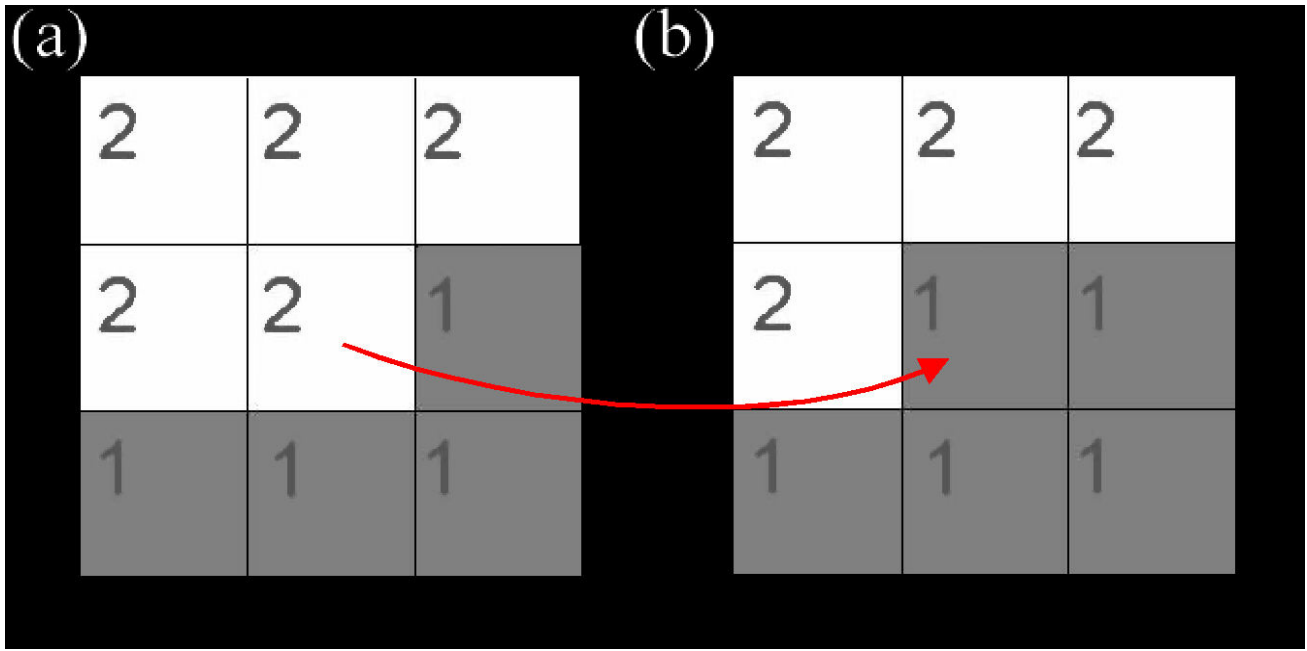

- **Figure 7**: Patch for the evaluation of pupal cam insertion: (a) patch with central pixel belonging to pupal chamber and (b) its replacement to gallery pixel. -

As an example, figure 8 show a slice showing chambers and pupal chambers obtained with the first separation algorithm where appears pupal chambers inserted in the galleries (a) while after the application of the second algorithm, the same part appears correctly labeled as part of the galleries.

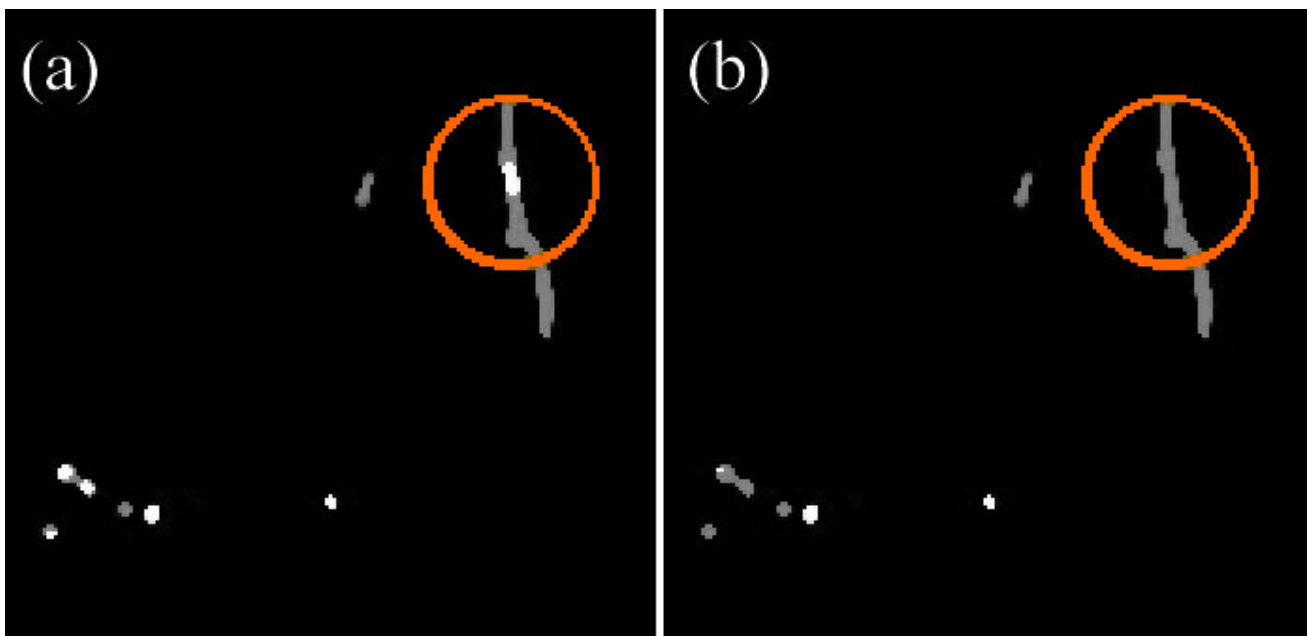

- **Figure 8**: Effect of the algorithm for detection of pupal chamber insertion: (a) original image showing a gallery with the insertion of the detected pupal chamber (red circle) for a given slice of the tomography and (b) the same region after the application of the algorithm. -

**Robustness of the method**

In order to highlight the ability of the proposed method to detect just object of the right characteristic size (i.e. the galleries and pupal chambers, whose characteristic size is determined by setting an adecuaque kernel), the set of algorithms was applied to a trunk which was mechanically sliced along its axial direction. Figure 9a shows a tomographic slice of this trunk, where the cuts appears as equispaced parallell strigth lines at the same angle. These lines could be confused with galleries by the algorithm. However, since the width of the cuts does not match the charactestic width of the galleries and pupal chambers, the distance map generated by the main algorithm show the slices atenuated while the galleries and pupal chambers appear more enhanced (see Fig. 9b). The application of the second and third thresholding algorithm to Fig 9b, correctly retrives the galleries, in grey, and the pupal chambers, in white (see Fig. 9c).

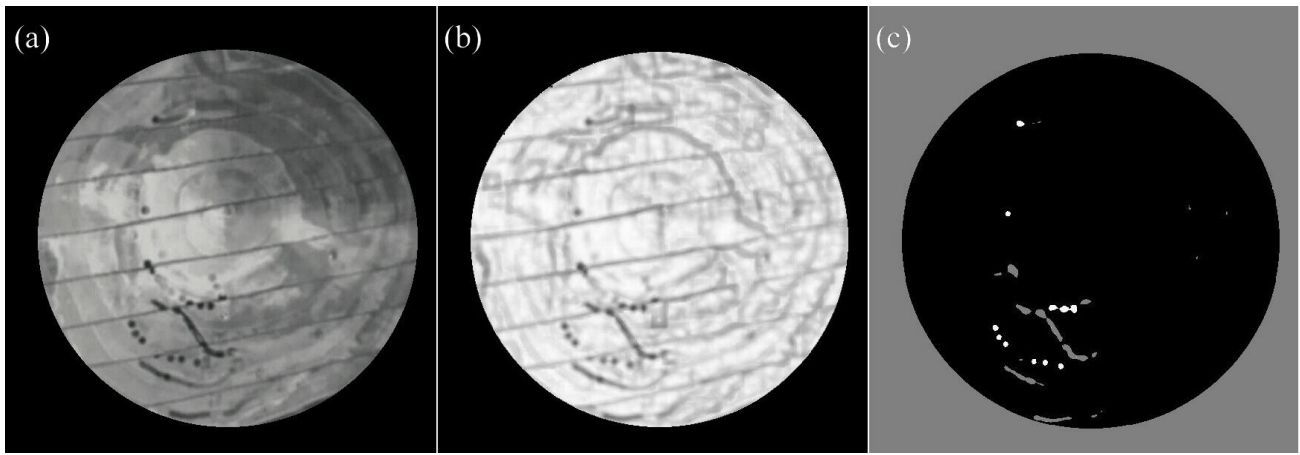

- **Figure 9**: Robustness of the detection algorithms: (a) tomographic slice of the mechanical sliced trunk; (b) corresponding distance map and (c) final detection thresholded image. -

**Final results**

Once each tomographic slice was processed with the set of algorithms, the pupal chambers and galleries can be easily visualized in order to be analyzed and quantified. Using any visualization software, the set of processed images can be join together in a bulk to generate a 3D representation of the three-dimensional structure. Figure 10 show to different points of view of the 3D structure showing the galleries and pupal chambers. The output of the algorithm could be used to automatically calculate the dimensions of the structure by means of a simple algorithm (not described in here). The following dimensions could be obtained:
- Galleries total volume: 26.12 $cm^3$
- Pupal chambers total volume: 7.27 $cm^3$
- Galleries total longitude: 215.00 cm
- Number of pupal chambers: 41

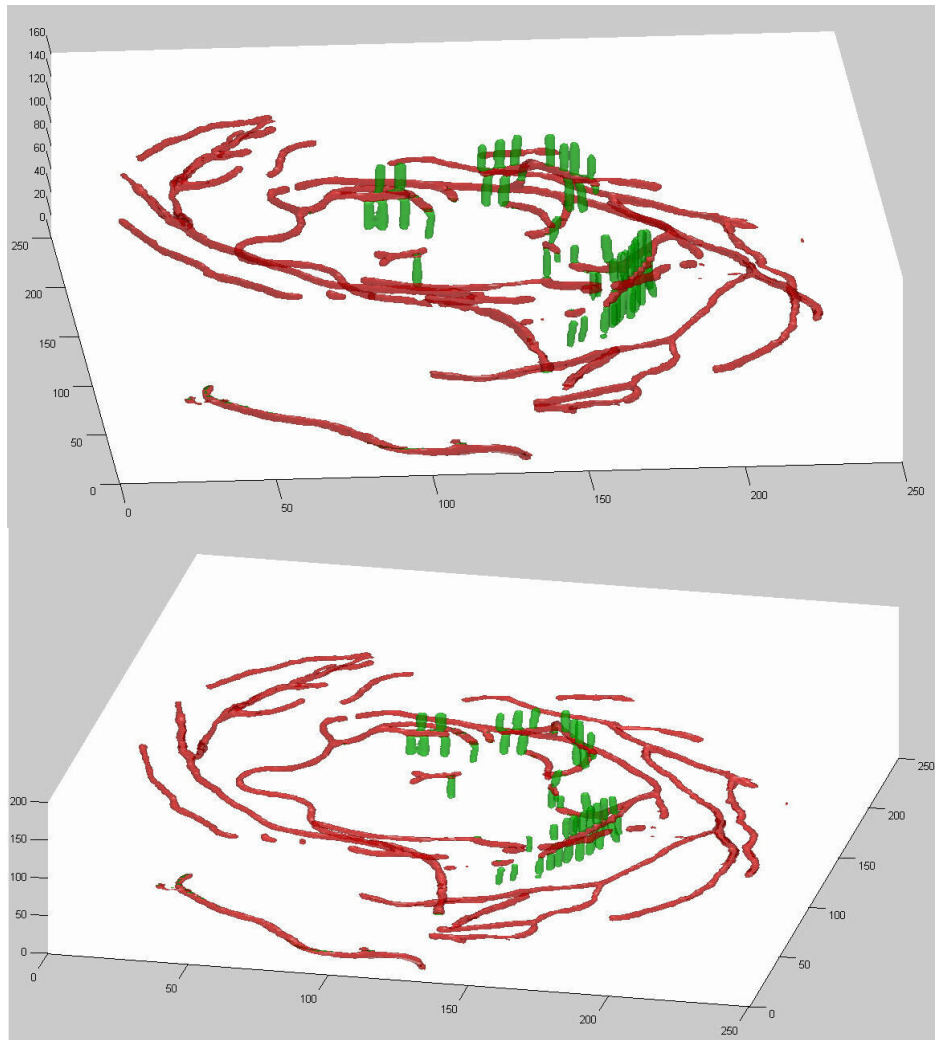

- **Figure 10**: Two different views of the 3D structure of galleries (in red) and pupal chambers (in green) obtained with the set of algorithms described in this work. -

**Conclusions**

Here we present a set of three algorithms to be able to isolate the images of galleries and pupal cameras in computerized axial tomografies of trunks affected by diffent typs of insects that bore galleries in the trees structure. The algorithms work concatenated. The first algorithm is the core of the procedure and basically it recognizes geometrical shape having a characteristic size in any direction by means of the definition of a certain abstract euclidean distance. The second algorithm is used to threshold the images provided by the second algorithm, while the third algorithm separates structures belonging to galleries from those belonging to the pupal chambers. We consider that the proposed methodology must find application in many areas of digital image processing and recognition. Future work will be aimed to find a set of algorithms that allow the automatic analysis of the topology of the galleries and pupal chambers structure, in order to be able to quantify additional parameters like number of bifurcations in the galleries, number of pupal chambers, curvature radiuses in the galeries, among others. In this manner, the analysis of several trunks could be used to obtain important biological information about the behaviour of the ambrosia beetle, allowing ths study of modulation factos of trunk damage, which in turn makes possible the implementation of measures of control of the plague.

## Acknowledgements

This work was realized thanks to the support of PICT-2019-00100. The authors also want to thanks Centro Medico Rossi for the tomographic images of the trunks used in this work.

## Author contribution

A.D. conceived and developed the algorithms. E.C. and C.C. provided the trunk tomographies. Y.C. processed the tomographies using the presented algorithms. Y.C. and E.C. analyzed the results. A.D. and E.C. wrote the first version of the manuscript. All the authors discussed the results and reviewed the manuscript.